\documentclass[11pt,a4paper]{article}
\usepackage{graphicx}
\usepackage{epstopdf}
\usepackage{epsfig,amssymb}
\usepackage[portuges,english]{babel}
\usepackage{amssymb}
\usepackage{amsmath}
\usepackage{amsfonts,a4wide}
\usepackage{tcolorbox}
\usepackage{enumerate}
\usepackage[latin1]{inputenc}
\usepackage{pgfplots}
\usepackage[linktocpage]{hyperref}

\usetikzlibrary{intersections}
\usetikzlibrary{patterns}
\DeclareGraphicsExtensions{.eps,.art,.ART,.ps,.jpg}

\newcommand{\beq}{\begin{equation}}
\newcommand{\eeq}{\end{equation}}
\newcommand{\be}{\begin{equation}}
\newcommand{\ee}{\end{equation}}
\newcommand{\beqa}{\begin{eqnarray}}
\newcommand{\eeqa}{\end{eqnarray}}
\newcommand{\beqar}{\begin{eqnarray*}}
\newcommand{\eeqar}{\end{eqnarray*}}
\newcommand{\bea}{\begin{eqnarray}}
\newcommand{\eea}{\end{eqnarray}}

\newcommand{\beg}{\begin{enumerate}}
\newcommand{\en}{\end{enumerate}}

\begin{document}

\begin{titlepage}
\hfill ~
\vskip 2cm
\begin{center}
{\LARGE \bf The ubiquity of black holes in modern physics}
\end{center}

\vspace{1cm}

\begin{center}
\Large Jorge V. Rocha

\vspace{0.5cm}
{\normalsize Departamento de Matemática, ISCTE -- Instituto Universitário de Lisboa,\\
Avenida das Forças Armadas, 1649-026 Lisboa, Portugal

\medskip
and
\medskip

CENTRA, Instituto Superior Técnico -- IST, Universidade de Lisboa -- UL,\\
Avenida Rovisco Pais 1, 1049-001 Lisboa, Portugal

\vspace*{0.5cm}

\texttt{jorge.miguel.rocha@iscte-iul.pt} }

\end{center}

\vspace{1cm}

\begin{abstract}
This is a translation of an article written originally in Portuguese for the journal {\it Gazeta de Física}, in a special edition celebrating the 2020 Nobel prize in physics. The text targets a  broad audience and focuses on the key ideas and developments in an entirely non-technical fashion, requiring from the reader only a keen interest in theoretical physics. The aim is to explain in an accessible way ---employing only basic and intuitive concepts---
the huge influence black holes have had in modern theoretical physics. Among the various topics covered, string theory and the AdS/CFT correspondence, extra dimensions, the quark-gluon plasma and holographic superconductors are highlighted.
\end{abstract}

\end{titlepage}


\tableofcontents


\vfill

\begin{minipage}[t]{0.55\textwidth}
~
\end{minipage}
\begin{minipage}[t]{0.45\textwidth}
\textit{\small
O sonho é ver as formas invisíveis\\
Da distância imprecisa, e, com sensíveis\\
Movimentos da esp'rança e da vontade,\\
Buscar na linha fria do horizonte\\
A árvore, a praia, a flor, a ave, a fonte --\\
Os beijos merecidos da Verdade.}
\end{minipage}

\begin{flushright}
\textit{\small Fernando Pessoa, from the poem ``Horizonte'' in ``Mensagem''}
\end{flushright}

\vspace{2cm}

\setlength{\unitlength}{1mm}

\section{The irresistible attraction of black holes\\{\sf \large Introduction}}

The most recent Nobel prize in physics was awarded in October of 2020 jointly to Roger Penrose, for the theoretical discovery of the inevitability of black holes, and to Reinhard Genzel and Andrea Ghez, for the observational confirmation of a supermassive black hole located at the center of our own galaxy~\cite{NobelPhysics202}.

Putting events in perspective, it should be noted that this recognition was bestowed more than one hundred years after the discovery of the first nontrivial solution of general relativity, corresponding to a static black hole in isolation. It also arrived more than 50 years after Penrose published the results that earned him the Nobel prize! Even the genesis of the term ``black hole'' dates back to that golden era of general relativity~\cite{Herdeiro:2018ldf}. However, in the last two decades black holes have shined under the spotlight.

Several factors contributed to this phenomenon. Among them, one must first highlight the long-awaited detection of gravitational waves by the LIGO and Virgo collaborations in 2015~\cite{LIGOScientific:2016aoc}, which brought along with it another type of waves ---those of excitement!---, and which was also awarded the Nobel prize in 2017. Secondly, the obtention of the first image of a black hole~\cite{EventHorizonTelescope:2019dse} also stands out as a milestone. Both of these occasions had an enormous impact in the media, which reflects its importance ---much like the confirmation of the supermassive black hole at the center of the Milky Way.

Black holes are extreme deformations of space and time, so drastic that they cause the appearance of what is called an {\it event horizon}. This horizon is an imaginary surface that encloses a region of spacetime from within which nothing can escape, not even light.

Having in mind the definition above, the astrophysical relevance of black holes in the cosmos is quite obvious. A lot more can be said about that topic, but it is not the main focus of this article. Much harder to understand is the importance of black holes for other subjects in physics, such as high energy physics or condensed matter. Yet, these intriguing objects play a central role in recent theoretical developments in multiple fields of modern physics. One might rightfully say that the black hole's irresistible attraction congregated around it many branches of physics that are apparently unrelated.

The rest of this article aims at giving a flavor of the fascinating connections between the science of black holes and other areas of physics. Namely, we will touch upon topics like extra dimensions, the quark-gluon plasma, or superconductivity, and their respective relations with black holes. All these elements find a place in the context of string theory, albeit the links between them are often far from being obvious. What allowed us to realize such connections exist? The main advancement that should be held accountable for this progress was a theoretical development of the end of the twentieth century, which was equally predominant in the recent ascent of black holes, known under the name of ``AdS/CFT correspondence'', or more simply ``holography'', in academic jargon. This field of scientific knowledge has grown tremendously, and it has become an independent and prominent subject of study. Nevertheless, it has its origins also in the context of string theory, to which we turn to next.

Before engaging with the subject, a brief note about the bibliography is in order. Since this article is not intended as a comprehensive review, the reference list is far from being complete, and many relevant works have been undeservingly left out. We direct the interested reader to the books [5, 6] and to the review articles [10, 12, 18, 25, 26] for further details.

\pagebreak

\section{Gravitating strings \\{\sf \large String theory and general relativity}}

String theory had its origins in the end of the '60s. Curiously, it was initially proposed as a potential candidate to describe the strong interaction, which appeared at the time as the greatest mystery in particle physics~\cite{Green:1987}. Meanwhile, as years passed by, it became clear that quantum chromodynamics, based on a quantum field theory framework ---and in that respect similar to the theory of electroweak interactions, already understood by that time---, could describe correctly the strong interactions. But string theory was not immediately abandoned, since it showed other virtues\dots

String theory is built over the simple and appealing idea that elementary particles and the forces that mediate them are nothing but different vibrational modes of sub-microscopic unidimensional objects~\cite{Green:1987,Zwiebach:2004tj}. In essence, the same happens with a violin string, which can produce several notes depending on how the string is allowed to vibrate, although the characteristic scales are completely distinct: the fundamental objects of string theory typically have sizes of the order of the Planck length, which is the scale at which gravitational, quantum and relativistic effects all contribute with equal relevance. To have an idea of the enormous disparity of scales, the proportion is roughly the same as the size of a proton relative to the extent of the Milky Way!

What, then, has string theory to offer that quantum chromodynamics hadn't solved yet? Besides its built-in ability to generate a multitude of particles from a single fundamental constituent, there is (at least) one other excellent reason to take such a theory seriously: it automatically incorporates the unification of all known forces of nature.

Let us contrast these attributes with those of the standard model of particle physics, formulated in the language of quantum field theories. The standard model is the upshot of more than 100 years of studies and offers a deep understanding of the behavior of elementary particles and their quantum interactions mediated by the electromagnetic, weak or strong forces. Nevertheless, this formalism embarrassingly leaves out the only remaining known force in nature: gravitation.

Just like photons (``particles of light'') can be understood as the mediators of the electromagnetic field, the gravitational field also has its own mediators: the {\it gravitons}. However, the methods used to ``quantize'' the electromagnetic, weak and strong interactions hopelessly fail when applied to the gravitational field.
This is where string theory comes to the rescue.
It turns out that the vibrational modes of fundamental strings reproduce not only the properties of the particles already included in the standard model, but also feature a particle with the correct behavior to play the role of the graviton. Moreover, the Einstein equations ---which govern all of general relativity--- magically emerge from string theory!

Hence, string theory accomplishes the long sought-after unification of fundamental forces. This feat is so much more incredible because, in the process, it provides a formalism to consistently tackle the gravitational interaction under the rules of quantum mechanics. String theory is therefore a theory of quantum gravity, something that Albert Einstein himself searched for for decades, unfortunately and uncharacteristically, without being successful.

\section{Strings wear black \\{\sf \large Black holes in string theory}\label{sec3}}

Since gravitation is encompassed by string theory, it is not surprising that black holes also make an appearance in its realm.

Astrophysical black holes are formed when a massive body is crushed ---under the action of gravity, and even opposing other forces--- into a sufficiently small region. Let us imagine that we take a ball and start compressing it in all directions equally. The critical value of the ball's radius for which a horizon appears is directly proportional to the mass of the sphere. For a ball with the mass of the Sun, this critical value, known as the Schwarzschild radius, is about 3 km. In other words, if the Sun were reduced to slightly less than the size of the Halley comet, it would become a black hole.

What is then the situation with strings?
The mass of a string grows (linearly) with its length, in such a way that each additional vibrational mode increases the associated mass. A simple calculation based on ``random walks'' shows that a sufficiently excited string typically has a smaller size than its own Schwarzschild radius. Such a configuration should naturally originate a black hole.\footnote{This argument ignores interactions between strings. Nevertheless, this effect tends to further compress the structure, so it does not invalidate the reasoning.} This is the key idea that was expressed in the correspondence principle between black holes and strings, during the '90~\cite{Susskind:1993ws,Horowitz:1996nw}.

Formally, a black hole in string theory can be considered as a solution of the equations that govern the {\it several} fields involved, while featuring a horizon. We stress the word ``several'', since string theory accommodates many other fields that frequently interact, in addition to the gravitational field. As a consequence, a black hole is not exclusively determined by the corresponding gravitational field --- it might have multiple associated charges. A simple example is helpful to support this claim: in the context of Einstein-Maxwell theory an electrically charged (and static) black hole is more compact than a neutral black hole with the same mass.

The previous paragraph suggests that there is a wide variety of black hole solutions in string theory. Yet, the framework has a structural characteristic that raises this multiplicity to a whole new level: basic consistency of the formalism demands spacetime to be ten dimensional.
This allows an even greater multitude of black holes in string theory, as we will see in the next section. For now, we shall content ourselves with noting a common trace shared by black holes in string theory: they can be infinitely extended in some directions~\cite{Horowitz:1991cd}. In such cases, they go under the name of black {\it branes}\footnote{The term ``brane'' stems from the word ``membrane.''}.

\subsection{A curse becomes a blessing\\{\sf \normalsize Extra dimensions}}

At first sight, the extra dimensions required by string theory seem to be inconvenient. To make contact with reality, all the additional dimensions have to be somehow hidden from our perception. The ways to achieve this can be classified into two big categories: compactification models and braneworld models. Each of these methods deserves a whole dedicated article and we will not delve much into none of them. Nevertheless, it is worth mentioning the key ideas on which they are based.

In compactification models the extra dimensions are curled up in sub-microscopic spaces. An illustrative analogy can be made with the help of an ant moving along the exterior surface of a hose: the insect can choose to move in any of the infinite possible directions, including preserving the alignment with the overall direction of the tube, or transversally (in which case it will not make much progress); but if we observe the hose from far away, we loose the notion of its thickness and it will be perceived as a one-dimensional object that can only be traversed in one direction. The existence of many possible viable ways to compactify space is a hallmark of these methods.

\begin{figure}[t]
\begin{center}
\includegraphics[width=\textwidth]{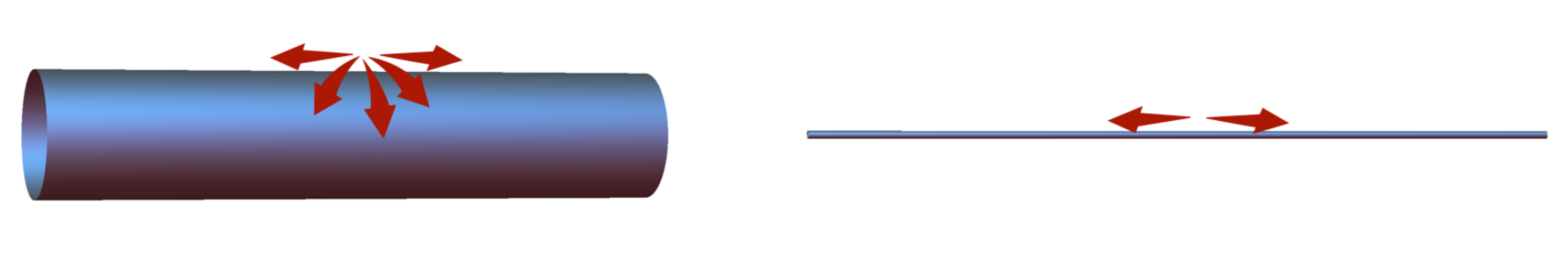}
\caption{A tube with its bidimensional surface (left) looks like a zero-thickness string when observed from sufficiently far away (right).}
\label{Tubes}
\end{center}
\end{figure}

\vfill
\pagebreak

The idea behind brane-world models is completely different. In those setups the constraints on the size of the extra dimensions are not so tight, but our physical reality is restricted to just three of the available spacial dimensions. The possibility that the dynamics of the brane-world in which we live could be related to the cosmological evolution of our universe is one of the most appealing aspects of these models.

In any case, extra dimensions have an interesting virtue: they greatly enlarge the configuration space. And this gives rise to an associated increase in the space of solutions, as we discuss next.

\subsection{More dimensions makes for more storage\\{\sf \normalsize Black holes with extra dimensions}}

In spite of being fascinating, black holes in general relativity are, with its presumed four dimensions, the simplest objects in the universe. In the absence of substancial matter in its vicinity or of perturbations caused, for instance, by other nearby black holes ---as was the case of the first detections from the LIGO-Virgo collaboration--- they are completely characterized by just two numbers: their mass and angular momentum. This claim is backed by certain uniqueness theorems~\cite{Chrusciel:2012jk}, and therefore it is an inevitable conclusion\dots~Unless the hypotheses of these theorems are evaded.

One such hypothesis concerns the number of spacetime dimensions. Namely, four dimensions are assumed. But, as we saw, string theory requires extra dimensions and this offers more possibilities, including the already mentioned existence of several black brane solutions. Yet, if we (theoretically) look for black holes {\it localized} in space, what do we find?

\begin{figure}[t]
\begin{center}
\includegraphics[width=\textwidth]{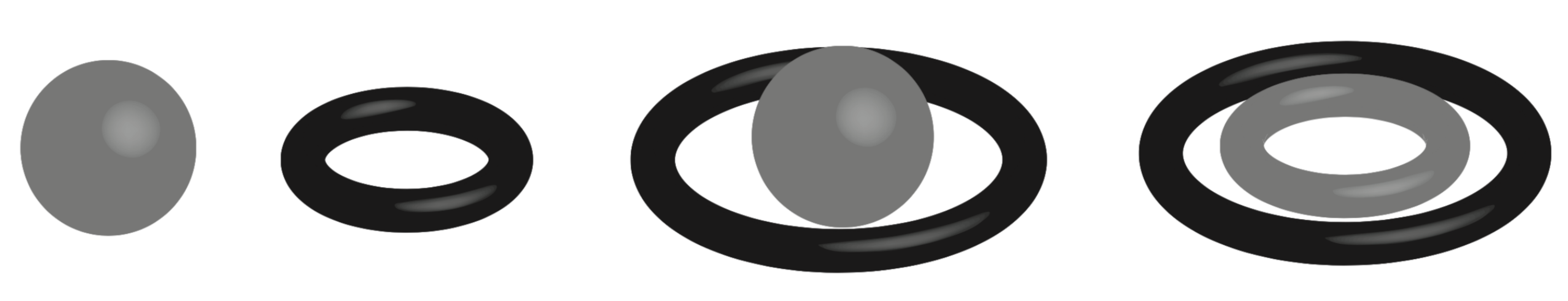}
\caption{Some of the known black holes in five dimensions, displaying distinct topologies of the event horizon. One of the four spatial dimensions is suppressed for illustration purposes. The color coding was chosen so that disconnected horizons can be easily distinguished, and has no particular meaning.}
\label{BHs_in_5D}
\end{center}
\end{figure}

Curiously, the uniqueness theorems are not applicable in more than four spacetime dimensions. This was explicitly demonstrated when, in 2001, the first black {\it ring} solution of Einstein's gravity theory in {\it five} dimensions was presented~\cite{Emparan:2001wn}. Since then many other solutions ---illustrated in figure~\ref{BHs_in_5D}--- have been discovered in 5D or higher dimensions~\cite{Emparan:2008eg}.

\subsection{Radiation from where it was least expected\\{\sf \normalsize Black hole thermodynamics}\label{BHthermodynamics}}

An important aspect of black holes is that they have an associated temperature, which is a surprising statement. It is surprising because it implies that it must radiate but, according to the classical definition of a black hole, nothing ---not even light--- can escape from inside a black hole.

\begin{figure}[t]
\begin{center}
\includegraphics[width=0.7\textwidth]{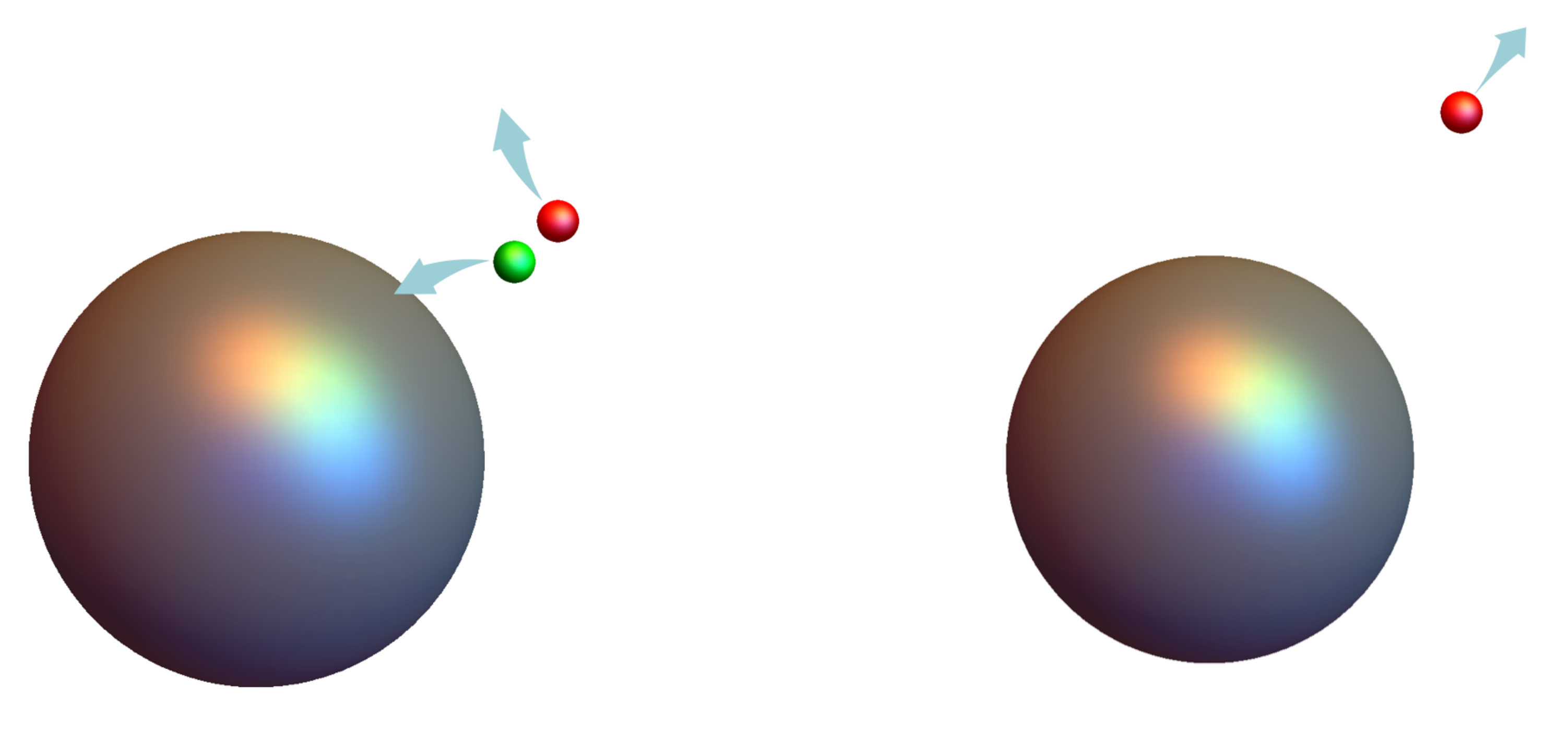}
\caption{Heuristic illustration of the Hawking effect. Every so often, quantum fluctuations spontaneously create a particle-antiparticle pair. When this occurs in a black hole neighborhood it can happen that one of the particles (carrying negative energy) falls through the horizon, while the other one escapes to infinity. In the process, the mass of the black hole is reduced.}
\label{HawkingRadiation}
\end{center}
\end{figure}

The apparent contradiction in the previous paragraph can only be clarified within the framework of quantum gravity. Perhaps the most intuitive way to understand this phenomenon is to recall that quantum fluctuations can momentarily cause particle-antiparticle pairs to randomly pop up. When this occurs close to a black hole, it might happen that one of the particles falls through the event horizon, while the other escapes to infinity (see figure~\ref{HawkingRadiation}). In order to conserve the total amount of energy in the system, the absorbed particle must carry with it some negative energy. The net result is that the black hole emits radiation and looses mass in the process.

This heuristic explanation for a black hole's quantum emissivity is firmly supported by a rigorous semiclassical calculation performed by Stephen Hawking in 1975~\cite{Hawking:1975vcx}. That analysis also revealed that the radiation emitted by a black hole ---the Hawking radiation--- has a characteristic blackbody spectrum, with a well defined temperature. This temperature is related with the mass of the black hole, with this relation usually being an inverse proportionality\footnote{Further ahead we will consider black holes living in other kinds of spaces, for which this relation is modified.}.

Historically, the path leading to this discovery was a bit less straightforward. In the first half of the '70s it became apparent that suggestive similarities exist between the laws that govern global properties of black holes and the laws of thermodynamics~\cite{Bardeen:1973gs}. This analogy can be promoted to a {\it quantitative} agreement if we are willing to assign the Hawking temperature to black holes, in addition to a corresponding entropy, which was identified by Jacob Bekenstein~\cite{Bekenstein:1973ur}. The correct entropy is proportional to the {\it area} of the event horizon. This is another curious feature of black holes (actually, of gravitational systems) ---usually the entropy grows with the volume of the object, instead of its surface area.

The assignment of an entropy to black holes raises an obvious question. In statistical mechanics, entropy provides a measure of the uncertainty about the macroscopic state of a system, simply because there exist several possible microstates compatible with the same course-grained description.
Then, what are the microstates of a black hole, accounting for its entropy? General relativity does not seem to offer any answer to this question. But a resolution can be found in the context of string theory. In fact, we saw in section~\ref{sec3} that, within this framework, a black hole can be described as a collection of highly-excited strings. What we did not mention before ---but is equally important for portraying a black brane--- is that there are other extended objects in string theory. This refers to the colloquially known {\it D-branes}~\cite{Polchinski:1995mt}, which, among other aspects, are entities on which the endpoints of fundamental strings can terminate.
The possibility of interpreting a black hole as a multitude of D-brane configurations connected by strings manifestly offers many more degrees of freedom than a bare spacetime geometry, which is the only viable description in general relativity. The correct identification of the entropy of black holes through the counting of such microstates\footnote{The microstate counting that reproduces the entropy of a black hole has been achieved only for certain classes of black holes with special properties that make the calculation feasible.} remains until today as one of the greatest successes of string theory~\cite{Strominger:1996sh}.

\vfill
\pagebreak

\section{The rise of holography\\{\sf \large The AdS/CFT correspondence}\label{Holography}}

It is impossible to talk about the successes of string theory without mentioning its most impactful offspring, which is certainly one of its pinnacles: the AdS/CFT correspondence.\footnote{Many excellent references devoted to the AdS/CFT correspondence can be found in the literature. The review article~\cite{Horowitz:2006ct} offers a very concise, clear and non-technical exposition.}

The AdS/CFT correspondence was a remarkable development in the realm of string theory in the late '90s, and it all began with an incredible ---but well-supported--- proposal by Juan Maldacena~\cite{Maldacena:1997re}.
This surprising correspondence postulates the full equivalence between two apparently unrelated theories! In such a case, the two theories are said to be {\it dual} to each other.
One of the protagonists is a quantum gravity theory ---one that can also be simplified to a classical theory of gravity in a certain limit--- performing in a very peculiar stage: anti-de Sitter space, or AdS for short, about which we will have more to say later.
The other actor is a quantum field theory (similar to quantum chromodynamics) in which gravitation is simply absent ---it was not even invited to the party! In the original studies this theory came equipped with a high degree of symmetry. In technical jargon, it is called a {\it conformal} field theory, from which the acronym CFT is derived. Nevertheless, nowadays we know that there are means of breaking this conformal symmetry in such a way as to obtain quantum field theories better adapted to describe nature.

Initially, the amazing duality put forward by Maldacena had a somewhat abstract character. But in a matter of a few months the correspondence was formulated in a precise fashion~\cite{Witten:1998qj,Gubser:1998bc}, giving it the status of an extremely useful theoretical tool, with enormous computational potential. Despite all this, the AdS/CFT correspondence still remains a conjecture: a rigorous proof of its veracity has proved hard to accomplish, but one should keep in mind that the evidence in its favor is overwhelming, both in quality as in quantity.

It is not the aim of this article to explain the many reasons behind our firm belief in the validity of this duality. To that end, several first-rate reviews on the subject already exist. Instead, we will content ourselves in highlighting some of its main properties and virtues for the remainder of the paper.

Let us start by tackling the most pressing question: ``How is it possible that two such different theories can be equivalent to each other?'' One of the most striking aspects about AdS/CFT that contributes to, at least, not reject this proposal immediately is that the two theories are not formulated in the same physical space; they do not even live in spaces with the same number of dimensions! The gravitational theory possesses more dimensions, and one of them is associated with the energy scale in the dual theory. The conformal symmetry of the field theory translates into the requirement that the ambient geometry for the gravitational theory is also highly symmetrical, but with one particularity: the curvature is negative (and constant throughout spacetime).

The feature just mentioned defines AdS space.\footnote{An example of an object whose surface has negative curvature, and approximately constant, is a saddle. This should be contrasted with a spherical surface, which also has constant curvature (but in that case it  is positive).} This geometry has a bizarre characteristic: although it is an unlimited space, light rays propagate out to infinity (and back) in a finite time. For all practical purposes, AdS behaves like a finite box. In a certain sense, the CFT can be thought of as living on the boundary of AdS. If we accept that the two theories are indeed equivalent, then every process occurring in the gravitational side must have a dual description in a space with one less dimension. Therefore, this is a {\it holographic} duality, one which explicitly implements the holographic principle enunciated by Gerard 't Hooft a few years before~\cite{tHooft:1993dmi}. Holography, which had been experimentally discovered more than 50 years earlier, thus became the center of the attention for hoards of theoretical physicists.

As we have alluded to above, the AdS/CFT conjecture posits the existence of an exact dictionary relating concepts and quantities on both sides of the equivalence. A significant amount of the efforts conducted in this area has been devoted precisely to putting together such a dictionary. One of the main items in this mapping has to do with coupling constants, which generically determine the strength of the interactions between the fundamental elements of each theory. It turns out that, in AdS/CFT, the coupling constants are inversely proportional to each other. This implies that when one of the sides is strongly coupled, the dual theory is weakly coupled, and vice versa. Hence, AdS/CFT is also an example of a strong-weak correspondence. This aspect is probably the greatest obstacle standing in the way of a mathematical proof of the equivalence. However, it is also its greatest virtue: if we face a difficult problem in a strongly coupled regime in one side of the duality, we can map it into an equivalent problem in a weakly coupled context, for which simple perturbative methods can be employed to make progress.

\subsection{Deconfining with black holes\\{\sf \normalsize Black holes and the quark-gluon plasma}\label{BHsAndQGP}}

\vspace{-1.2cm}
\hspace{6.6cm}\footnote{This title may, or may not, be inspired by the covid-19 pandemics.}

\vspace{0.8cm}
\noindent
In the vast conceptual scenario of the AdS/CFT correspondence, black holes play a crucial role, and they fit quite naturally in the gravitational side of the duality. We certainly can consider black holes in AdS and we know how to describe them. What is, then, the holographic representation of a black hole?

\begin{figure}[t]
\begin{center}
\includegraphics[width=0.7\textwidth]{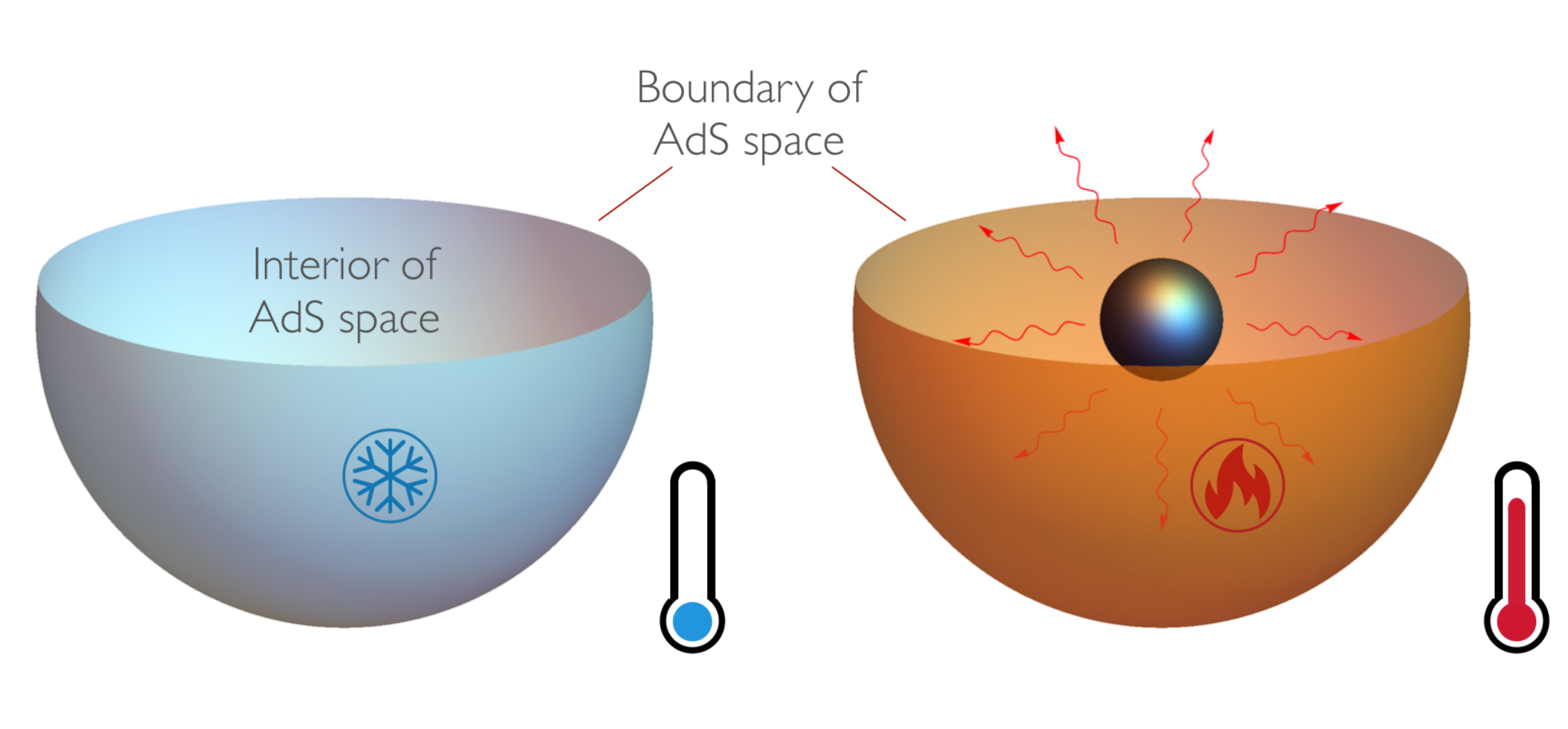}
\caption{Representation of AdS space as a finite box (and highly symmetric). The dual field theory may be considered, in a certain sense, to live on the boundary of this box. In the absence of any matter or black holes, the holographic dual to which it is associated is in a cold state (left). Inserting a black hole in AdS increases the temperature of the thermal state to which it corresponds (right).}
\label{AdSspace}
\end{center}
\end{figure}

To answer this question it is helpful to recall one of the lessons from section~\ref{BHthermodynamics}: black holes have an associated temperature. For the relevant AdS black holes, this temperature increases as their mass grows.
Typically ---but not always--- we do not have the option of populating the dual field theory with black holes. Nevertheless, it includes another kind of states with a well-defined temperature: a gas formed out of elementary particles in thermal equilibrium. The relation we seek can now be uncovered. A black hole in AdS with a given Hawking temperature is the holographic dual of such a gas at exactly the same temperature. We thus conclude that by inserting a black hole in AdS space we are effectively warming up the dual theory, and the bigger the black hole is, the hotter the corresponding state will be (see figure~\ref{AdSspace}).\footnote{Incidentally, the fact that the ambient space containing the black hole behaves asymptotically like AdS is crucial for the agreement between the entropies computed from the field theory side and from the gravitational dual.}

One of the most impressive checks on the validity of the AdS/CFT correspondence concerns the identification between phase transitions in the two dual theories. In fact, an {\it isolated} black hole in AdS is not in thermal equilibrium, since it emits Hawking radiation (thereby shrinking ever so slightly). Naturally, what occurs is that the full system evolves until the black hole and the gas of surrounding particles remain at the same temperature, in a steady state. Naively, one might think that such a configuration at any desired temperature could exist. However, this is not the case. Below a certain temperature (related to the curvature of AdS) the thermodynamically preferred state simply does not include a black hole, only a gas of particles at that temperature.This Hawking-Page transition, as it became known, is yet another surprise that AdS space had in store for us, one that had been discovered back in the '80s~\cite{Hawking:1982dh}.

How is this phase transition reproduced in the dual quantum field theory? In that framework, the only thing we have at our disposal are the fundamental constituents of the theory. It is helpful to borrow some intuition from quantum chromodynamics, where the elementary particles are identified as quarks and gluons. However, those degrees of freedom become apparent only at high temperatures, giving rise to the quark-gluon plasma. At low temperatures the elementary particles are confined, without ever making a solo appearance. The different possible combinations produce composites which we call hadrons, such as protons and neutrons. Hence, we also find a phase transition in this context ---one that has been dubbed the confinement-deconfinement transition. The fact that such a phase transition actually occurs in the holographic dual to AdS, and in such a way that the entropies computed from both sides of the correspondence agree, was uncovered by Edward Witten in a seminal article that kickstarted AdS/CFT~\cite{Witten:1998zw}.

\begin{figure}[t]
\begin{center}
\includegraphics[width=\textwidth]{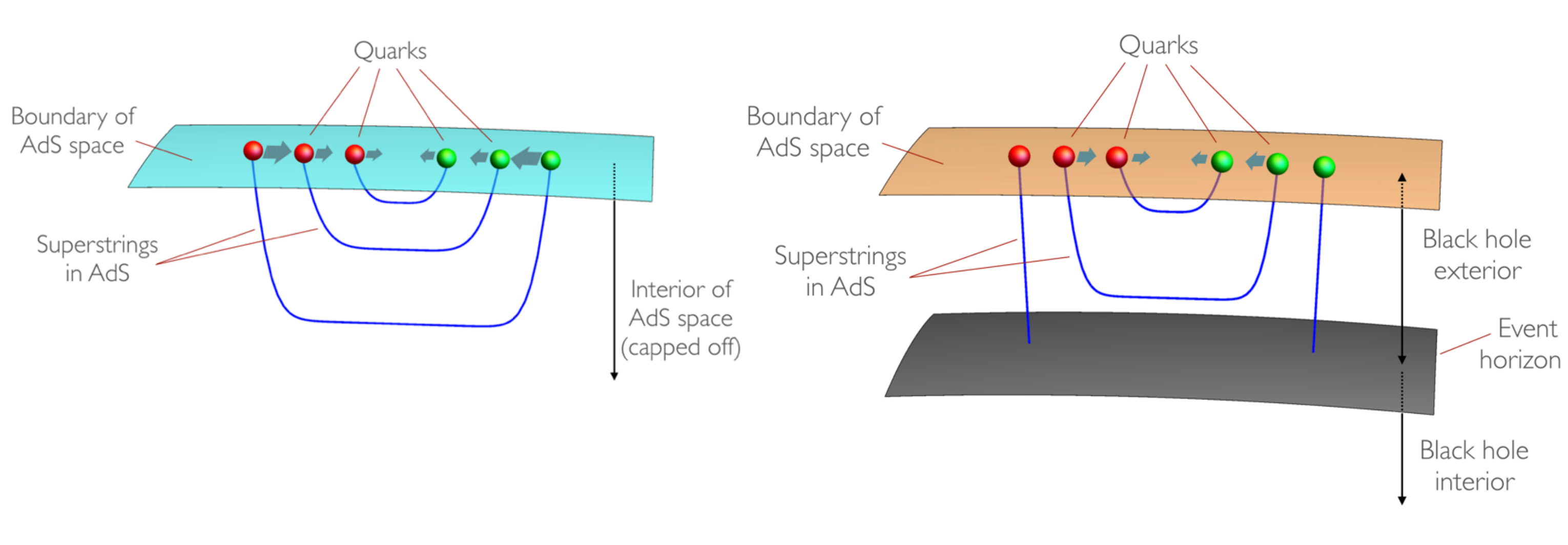}
\vspace{-0.7cm}
\caption{The confinement-deconfinement transition from the holographic point of view, focusing on quark-antiquark interactions. The quarks, constrained to move on the boundary of AdS, are connected by strings that fall into the interior of the dual space. In the absence of a horizon, the restitutive force that attracts the pair of quarks grows as the particles move apart and the string stretches. This low-temperature phase is illustrated in the left panel. In a high temperature phase, corresponding to the presence of a black hole in AdS, the strings can break up as shown in the right-hand panel. As a consequence, the quarks can then move freely, i.e., they cease to be confined.}
\label{DeconfinementTransition}
\end{center}
\end{figure}

There exists an intuitive argument that provides a graphical interpretation of this identification between the Hawking-Page transition and the confinement-deconfinement transition. It is based on the holographic description of the interaction between quark-antiquark pairs ---forming the so-called mesons---, which can be thought of as the endpoints of strings that ``hang'' from the AdS boundary. The fact that the strings have an intrinsic tension explains why the quarks are mutually attracted, and the restitutive force grows with the distance between the quarks. This description is appropriate when the system is at low temperatures. According to what was discussed previously, this occurs when there is no black hole in the interior of AdS, corresponding to the left panel in figure~\ref{DeconfinementTransition}.\footnote{In fact, for technical reasons the space that must be considered is not simply empty AdS, but instead a certain modification of it. The relevant aspect is that this altered space has exactly the same asymptotic behavior as AdS with a black hole, but without featuring any event horizon~\cite{Mateos:2007ay}.} As we increase the temperature by inserting a black hole in AdS, as illustrated in the right panel of figure~\ref{DeconfinementTransition}, the string that connects the quark-antiquark pair falls through the event horizon if it becomes too long. Consequently, the string snaps and the quarks become free.

\vfill

\subsection{Black holes offer no resistance\\{\sf \normalsize Superconductivity from holography}}

Another application of AdS/CFT in which black holes play a central role is concerned with the so-called holographic superconductors. An important phase transition also occurs in this setting ---one that can be understood in much the same way as what was described in section~\ref{BHsAndQGP}.

In order to somehow capture the essence of superconductors, we must introduce electric fields. What kind of modifications does figure~\ref{DeconfinementTransition} suffer when we consider AdS black holes that are electrically charged? Keeping this purpose in mind, adding charge to the black hole has a significant effect only if there exists some other field (or its quantum description in terms of particles) to interact with, through their combined electrodynamics. In this context, the simplest possible additional element is provided by an electrically charged scalar field.

\begin{figure}[t]
\begin{center}
\includegraphics[width=0.9\textwidth]{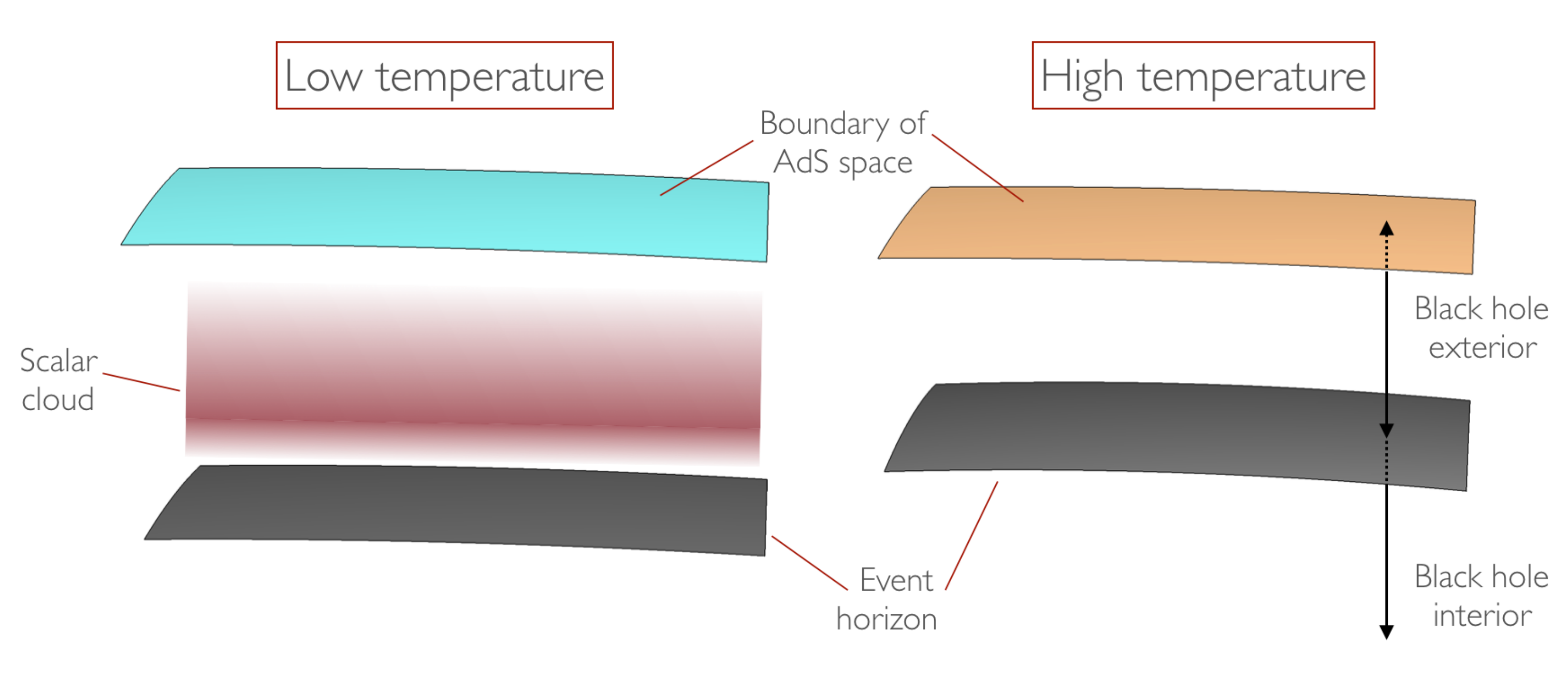}
\caption{Dual description of the phase transition occurring in holographic superconductors. At low temperatures (left) the event horizon of an electrically charged black hole sits far away from the boundary of AdS. This allows the formation of a cloud of charged scalar field, which is the dual representation of the condensate responsible for superconductivity. At high temperatures (right) there is less room between the horizon and the AdS boundary, and the scalar cloud does not fit in that region. Without the presence of a condensate, the conductivity drops abruptly.}
\label{HoloSuperconductor}
\end{center}
\end{figure}

A detailed discussion about holographic superconductors is a somewhat technical subject, but the key idea can be understood with minimal use of advanced physics~\cite{Horowitz:2010gk}. The electric field of the black hole can spontaneously create pairs of particles with opposite charges, by virtue of the well-known Schwinger effect. The frequency of occurrence of this quantum phenomenon increases with the intensity of the electric field, so it is predominant near the horizon. But the effect is also more relevant when the black hole's temperature is lower. The reason behind this is also connected with the strength of the electric field: colder black holes are smaller, and in that case the horizon sits closer to the origin of the electric field.
In any case, since the Schwinger particles have opposite charges, one of them is attracted to the black hole while the other is repelled. Given that AdS behaves like a finite box, the repelled particle cannot move away indefinitely. Instead, it will wander until it finds a stable position. The repetition of this sequence causes the accumulation of a cloud of charged scalar particles in the black hole's neighborhood (see left panel in figure~\ref{HoloSuperconductor}).

The condensation of a scalar cloud around a black hole happens only when the latter is sufficiently cold. At higher temperatures the event horizon covers a larger volume and the situation changes. On the one hand, the intensity of the electric field at the surface of the black hole is decreased. On the other hand, there is less room left in the black hole exterior and at a certain point the characteristic Compton wavelength of the scalar field no longer fits there, in contrast with what happened at lower temperatures (see right hand panel of figure~\ref{HoloSuperconductor}).

The charged scalar cloud that forms at lower temperatures is the holographic representation of the condensate of electron pairs with opposite spins, which are responsible for the phenomenon of superconductivity ---the Cooper pairs. Their presence indicates that the conductivity in such a phase is amplified and this prediction is indeed backed up by computations performed in the gravitational dual theory. In the context of AdS/CFT, both the expectation value of the condensate and the conductivity can be extracted explicitly as a function of the temperature of the system. The upshot is that below a certain critical temperature, the condensate acquires a non-vanishing expectation value, with a qualitative behavior similar to that experimentally observed in real world superconductors. This comes hand-in-hand with a formally infinite conductivity. Above the critical temperature the condensate disappears and the holographic system displays the usual electric resistivity.

So far, holographic models like the one just described can reproduce the behavior of real world superconductors only in a qualitatively correct manner. Nonetheless, the ability to explain superconducting phase transitions using black hole physics is remarkable. Even more so, if it is taken into account that other more conventional methods are not applicable to superconductors with high critical temperatures, usually. Such materials typically form in a strongly coupled regime and, as discussed in the end of section~\ref{Holography}, AdS/CFT is a particularly well adapted tool to employ under those conditions.

\section{In search of new horizons\\{\sf \large Conclusion}}

The ubiquity of black holes in modern physics is unquestionable. The advent of the AdS/CFT correspondence largely contributed to this state of affairs, as we have tried to demonstrate.

It is not at all surprising that black holes play a central role in theoretical physics. Being objects where essentially quantum mechanical phenomena can compete with purely classical gravitational effects, they form the ideal theoretical laboratory to unravel the mysteries of quantum gravity --- the current chimera in physics.

On the other hand, the influence that black holes have had on fields so loosely related, such as plasma physics or condensed matter, was, at least, unexpected.

Many other interesting and relevant subjects were not covered in this article, and the topics discussed represent just a sample of the extent to which the study of black holes contributed to broaden the horizons of scientific knowledge.

Thus, we owe much of the recent advances in theoretical physics to black holes. But who could have guessed that precisely these intriguing objects ---from where not even light can escape--- would enlighten this endeavor of human civilization?


\bigskip
\bigskip
\subsection*{Acknowledgements}

I am grateful to Antonia Frassino, Carlos Herdeiro and Roberto Emparan for helpful comments on the manuscript. I also thank Carlos Herdeiro and José Lemos, as editors of {Gazeta de F\'isica}, for the kind invitation to write up the article that served as the basis for this translation.
The author acknowledges financial support from {\it Funda\c{c}\~ao para a Ci\^encia e a Tecnologia} (FCT) through projects CERN/FIS-PAR/0023/2019 and UIDB/00099/2020.



\end{document}